\documentclass[12pt]{article}
\usepackage{amsmath,latexsym,amssymb,array}
\def\_{\hspace{-17pt}.}

\title{Geometrical Incompatibility \\ of Quantum Measurements}

\author{Georgy Parfionov \footnote{e-mail: your@GP5574.spb.edu} ,\;
Yury Romashev \footnote{e-mail: romashev@JR1822.spb.edu}\\ \small
A.A.Friedmann Laboratory for Theoretical Physics\\
\small St.Petersburg E--F University, Griboyedova 30--32\\
\small 191023, St.Petersburg, Russia }
\date{}
\begin{document}
\maketitle

\begin{abstract}
The problem of quantization of general relativity is considered in
the framework of noncommutative differential geometry. Operator
analogues for interval, scalar curvature, values of the Einstein
tensor are proposed. Quantum measurements of these observables
lead to a paradox:  different procedures of measurements can
supply non equivalent geometrical pictures of space-time. A
concrete example of such situation is provided.
\end{abstract}

\section{\_ The problem of
measurement}

The classical model of gravitation based on the Einstein equations
for the components of the metric tensor $g_{\mu \nu}$ does not
admit any measurement procedure for gravitational fields that
would be satisfactory from the quantum point of view. The
difficulties arise due to nonlinearity of the Einstein equations
and manifest themselves in measurements of field averages in small
domains of space-time \cite{Hawking}.

An analysis of the motion of test particles shows \cite{Infeld}
that there exists the lower bound $R=MG/c^2$ for the size of a
particle of mass $ M $. The derivation of this estimate is
essentially based on the nonlinear character of Einstein's
equations. On a whole, none of the known measurement procedures
yields satisfactory results if the linear dimensions of the domain
in which the measurements are made are less than $ L=\sqrt{Gh/c^3}
\approx{4\cdot{10^{-33}}}cm $ \cite{Regge}.  The microscale
parameter $L$ is the structural constant  in the Wheeler---De Witt
theory of quantum gravitation \cite{DeWitt,Wheeler}.

Thus, the microscale structure of space-time is in principle
indefinable because of the absence of the corresponding
measurement procedure. This fact calls in question the
applicability of the differential-geometric models of space-time
to microscales. In this connection, several attempts were made to
study more general objects than manifolds  \cite{Heller}, or even
to refuse from the concept of the point as an idealization of an
event in space-time  \cite{Geroch,ParZap}.  Among these attempts
one approach to quantization of general relativity first suggested
by Geroch  \cite{Geroch} seems to be the most promising. It uses a
reformulation of the classical theory in which the events of
space-time play essentially no role. This approach is based on the
well-known fact \cite{Kobayashi} that the main objects of
differential geometry can be formulated in a purely algebraic way
without any reference to the space-time continuum.

For example, a smooth vector field $ v $ on a manifold $ M $ can
be represented as a derivation on the algebra ${\cal A}(M)$ of
smooth functions on $ M $, that is, as a linear mapping $ v:{\cal
A}(M)\to{\cal A}(M) $ for which the Leibniz rule \mbox{$v(a
b)=v(a)b+av(b)$} holds true.

Thus, the vector field $ v $ can be considered as a purely
algebraic object. Tensor fields, connection, curvature and other
constructions used in general relativity can be obtained
analogously in the same algebraic way. Using these fact Geroch
suggested to take as the basic object of the theory not the
algebra ${\cal A}(M)$ of smooth functions on a manifold, but an
arbitrary commutative algebra ${\cal A}$ and then to construct the
differential geometry generated by ${\cal A}$ without any
reference to the underlying manifold, thus "smearing out events"
i.e. points of the manifold.

However, an analysis shows \cite{ParZap,Violette} that the
substitution of ${\cal A}(M)$ as the basic object of the theory by
a commutative algebra does not really enable one to "smear out
events". The fact is that any semi-simple commutative algebra
${\cal A}$ is canonically isomorphic to an algebra of functions on
a set $ M $ (the set of one-dimensional representations of ${\cal
A}$). Moreover, $ M $ assumes the topology (called the Gel'fand
topology) \cite{Milnor,Loomis} generated by the algebra ${\cal
A}$. It is essential here that if ${\cal A}={\cal A}(M)$, then the
algebraic structure of ${\cal A}$ permits to recover the set $ M
$, its topology and its differential structures.

It is clear that if the basic algebra ${\cal A}$ is commutative,
the proposed algebraization of space-time does not resolve the
problems of general relativity on microscales and the related
problems of measurement. The situation  completely changes if the
basic algebra ${\cal A}$ is noncommutative. In this case the
points of the underlying manifold cannot be recovered because the
noncommutative algebras cannot be represented functionally. In
\cite{ParZap} it is shown that all necessary geometric objects,
including the Einstein tensor can be obtained from a
noncommutative basic algebra.

\section{\_ Global geometry and quantization}

Omitting some details, we quote here the main constructions
proposed in \cite{ParZap} as well as an explicit expression for
the covariant derivative for the case of the Levi-Civita
connection. The basic object of a global geometry is an algebra
${\cal A}$. It should be emphasized that ${\cal A}$ is not
supposed to be commutative and can be interpreted as an algebra of
quantum mechanical observables. A derivation of the algebra ${\cal
A}$ is a mapping $ v:{\cal A}\to{\cal A} $ with the following
properties: \[v(a+b)=v(a)+v(b)
\]\[ v(a b)=v(a)b+av(b)\] The set of derivations $V$ is a Lie algebra
with the Lie bracket $[uv]=uv-vu$. If ${\cal A}={\cal A}(M)$, then
$V$ coincides with the Lie algebra of vector fields on $ M $. It
is natural to call the elements of $ V $ vectors. For $a\in Z$
(the center of the algebra $A$) the product $a v$ can be defined:
$(a v)(b)=a(v(b))$. The set $V^\ast$ of $Z$-homogeneous forms on
$V$ is the space (module) of covectors. There is the canonical
coupling between $V$ and $V^\ast$: $<f,g>=f(g)$ for $f\in V^\ast$,
$g\in V$.

In the case of manifolds $V^\ast$ is the space of covector fields
(differential 1~-~forms). The metric is introduced by a invertible
linear mapping $G: V\to V^\ast$ such that $g(u,v)=<Gu,v>$ is a
symmetric bilinear form. Then one can introduce:
\begin{enumerate}
\item {\bf The covariant derivation:} (Coszul's formula) for
$x,y\in V$, $z\in V^\ast$
\[
<z,\nabla_y x> =
\frac{1}{2}\{[G^{-1}z](g(y,x)+y(<z,x>)-x(<z,y>))+\]
\[+g(y,[x,G^{-1}z]) + g(x,[y,G^{-1}z]) - <z.[y,x]>\}
\]
Thus, $\nabla_y:x\mapsto \nabla_yx$ is the mapping of $V$ into
itself such that
\[
\nabla_y(ax)=y(a)\cdot x + a\cdot\nabla_yx
\]
\[
\nabla_{ay+bz}=a\nabla_y + b\nabla_z
\]
for $a,b\in Z$ and $x,y,z\in V$, and the torsion $T(x,y)=\nabla_xy
- \nabla_yx - [x,y]$ equals 0.

\item {\bf The  Riemann tensor} for $x,y,z\in V$, $w\in V^\ast$ \[
R(x,y,z,w) = <w,([\nabla_x,\nabla_y] - \nabla_{[x,y]})z> \] that
satisfies the Bianchi equality.

\item {\bf The contraction}, i.e. a linear form ${\rm
Ctr}:L(V,V^\ast)\to A$, where $L(V,V^\ast)$ is the space of
bilinear forms on $V\times V^\ast$, such as

\[{\rm Ctr}(K)={\rm tr}{\cal K}\]

 for the form $K(u,v)=<u,{\cal K}v>$, $u\in V^\ast$, $v\in V$.

\item {\bf The Ricci tensor}
\[
{\rm Ric}(x,z)={\rm Ctr}(K_{xz})
\]
where $K_{xz}(y,w)=R(x,y,z,w)$ \item {\bf The scalar curvature}
\[
r={\rm Ctr}(L)
\]
where $L(x,y)={\rm Ric}(x,Gy)$ and, at last, \item {\bf The
Einstein tensor:}
\[
{\rm Ric}-\frac{1}{2}gr.
\]
\end{enumerate} The global character of introducing of tensors
as elements of an abstract tensor algebra permits to suggest a
natural modification of the standard scheme of canonical
quantization i.e. representation of observables by operators. In
the given situation the values of tensors can be represented by
operators. The operator representation of scalar curvature is of
particular interest, because it is proportional to the contraction
of the energy-momentum tensor in the classical situation.A more
complicated example is the canonical quantization of the metric
tensor $g$. For $v\in V $ $a=g(v,v)$ is an element of the algebra
$\cal A$ and for any $v$ one can consider its operator
representation.

As a matter of fact, the representation of tensor values can be
reduced to the investigation of representations of the basic
algebra A and their relations with the differential-metric
structure of the three $({\cal A},V,g)$.

\section{\_ Spatialization procedure}

If the basic algebra $\cal A$ is the algebra of smooth functions
on a manifold $M$, the latter can be recovered by $\cal A$,
because all its irreducible representations are one-dimensional.
The one-dimensional representations form a continuous series
(Gel'fand representation) that can be "numbered" by the points of
the underlying manifold $M$, and thus we remain in the classical
situation.

The proposed scheme of canonical quantization has sense if the
basic algebra $\cal A$ is noncommutative. Hence it cannot be
represented as an algebra of numeric functions on a manifold,
because the dimensionality of at least one of its irreducible
representations should be more than one. Nevertheless, as we have
seen, all the main geometric objects can be constructed even in
this case. Thus,one obtains a geometry without points. This fact
should not provoke objections if the elements of $\cal A$ as
observables do not commute. That can be considered as an
expression of the complementarity principle. However, if in a
noncommutative algebra $\cal A$ one singles out a set of commuting
elements (simultaneously measurable observables), then the
subalgebra $\cal B$ generated by them will be commutative and
therefore can be represented by an algebra of functions on a
topological space $M$.

That enables one to consider $M$ as a space of events that can be
determined by a choice of commuting variables (generators of the
subalgebra $\cal B$). The procedure of a choice of a commutative
subalgebra and its geometric realization we call {\it
spatialization}. Thus, any choice of commuting variables gives a
certain geometric picture at the classical level. It is expedient
to use maximal commutative subalgebras for spatialization.

Note that under some additional conditions the spatialization of a
commutative subalgebra $\cal B$ of $\cal A$ is supplied not only
with a topology on the Gel'fand space $M({\cal B})$, but also with
a differential structure induced by the triple $({\cal A},V,g)$.
However, if the basic algebra $\cal A$ has two commutative, but
not commuting with each other subalgebras, the latter can generate
nonisomorphic geometries.

We call two spatializations do not commute if they are induced by
noncommutative subalgebras. Note that the corresponding geometries
even if they are isomorphic cannot be observed in one experiment.
It would be natural to ask, whether one and the same algebra of
observables can give rise to nonisomorphic spatializations. In
other words, does the quantum concept of measurement admit {\it
plurality of observed geometric pictures}?

\section{\_ An example}

Let is show that starting from a noncommutative algebra $\cal A$
one can obtain at least two topologically different
spatializations. Let $\cal A$ be the algebra of $2 \times
2$-matrices with the elements $a_{ik}=a_{ik}(x,y)$, $x,y \in R$,
that are smooth functions of two real variables. Let us consider
two subsets $\cal P$, $\cal Q$ of $\cal A$ defined by the
following relations:

\medskip
\noindent {\bf The set} $\cal P$:
\[ \partial_ya_{ik}=0,\quad a_{12}=a_{21}=0,
\]
\[ a_{11}(x+2\pi,y)=a_{11}(x,y),\quad a_{22}(x+2\pi,y)=a_{22}(x,y)
\]
{\bf The set} $\cal Q$:
\[ \partial_xa_{ik}=0,\quad a_{11}=a_{22},\quad a_{21}=a_{12}
\]
Any of the subsets $\cal P$ and $\cal Q$ is a maximal commutative
subalgebra of $\cal A$, however these subalgebras  do not commute.
For example, the matrices
\[ p=\left[\begin{array}{rr}
1&0\cr 0&-1
\end{array}\right] \qquad
q=\left[\begin{array}{rr} 0&1\cr 1&0
\end{array}\right]
\]
do not commute and $p \in {\cal P}$, $q \in {\cal Q}$. The
spatialization of the subalgebra $\cal P$ gives a disconnected sum
of two circles $S_1\cup S_2$. That corresponds to two continuous
series of one-dimensional representations
\[ \pi^1_s(p) = p_{11}(s,y),\quad
\pi^2_t(p) = p_{22}(t,y) \] where $s \in S_1$, $t \in S_2$. At the
same time analogous calculations for the subalgebra $Q$ yield
quite  different results. The spatialization of $Q$ gives  a
disconnected  sum of two straight lines $ R_1\cup R_2$ . In this
case one also obtains two continuous series of one-dimensional
representations
\[ \rho^1_u(q)=q_{11}(x,u)+q_{12}(x,u),\quad
\rho^2_v(q)=q_{11}(x,v)-q_{21}(x,v) \] where $u \in R_1$, $v \in
R_2$. It is plain that if two spatializations are topologically
different, all the more they cannot be isomorphic on a more
delicate differential-geometric level.

\medskip

\paragraph{Acknowledgments.} We express our gratitude to A.A. Grib
and R.R. Zapatrin for useful discussions.

\end{document}